\documentclass[pdftex,twocolumn,epjc3]{svjour3}          

\RequirePackage[T1]{fontenc}

\smartqed  
\RequirePackage{multirow}
\RequirePackage{graphicx}
\RequirePackage{mathptmx}      
\RequirePackage{flushend}
\RequirePackage[numbers,sort&compress]{natbib}
\RequirePackage[colorlinks,citecolor=blue,urlcolor=blue,linkcolor=blue]{hyperref}

\usepackage{tabularx}
\usepackage{graphicx}  
\usepackage{dcolumn}   
\usepackage{bm}        
\usepackage{amssymb}   
\usepackage{feynmf}
\usepackage{cancel}
\usepackage{hyperref}
\usepackage{slashed}
\usepackage{multirow}
\usepackage{color}
\usepackage{array}
\usepackage{subcaption}
\usepackage{amsmath}
\usepackage{booktabs}
\usepackage{multirow}
\usepackage{xspace}
\usepackage{lineno}
\usepackage{float}
\usepackage{dblfloatfix}

\newcommand*{\mum}{\textmu m\xspace}

\newcommand{\scd}{COCOA}

\begin{document}

\title{Configurable calorimeter simulation for AI applications}

\author{
    Francesco Armando Di Bello \thanksref{gen_add}
    \and Anton Charkin-Gorbulin \thanksref{lxm_add}
    \and Kyle Cranmer  \thanksref{nyu_add,uwm_add}
    \and Etienne Dreyer \thanksref{wis_add, e3}
    \and Sanmay Ganguly \thanksref{tok_add, e1}
    \and Eilam Gross \thanksref{wis_add}
    \and Lukas Heinrich \thanksref{tum_add}
    \and Lorenzo Santi \thanksref{rom_add}
    \and Marumi Kado \thanksref{mpi_add, rom_add}
    \and Nilotpal Kakati \thanksref{wis_add}
    \and Patrick Rieck \thanksref{nyu_add, e2}
    \and Matteo Tusoni \thanksref{rom_add}
}

\thankstext{e1}{e-mail: sanmay.ganguly@cern.ch}
\thankstext{e2}{e-mail: patrick.rieck@cern.ch}
\thankstext{e3}{e-mail: etienne.dreyer@weizmann.ac.il}

\institute{
    INFN and University of Genova \label{gen_add}
    \and University of Luxembourg \label{lxm_add}
    \and Weizmann Institue of Science \label{wis_add}
    \and New York University \label{nyu_add}
    \and University of Wisconsin–Madison \label{uwm_add}
    \and ICEPP, University of Tokyo \label{tok_add}
    \and Technical University of Munich \label{tum_add}
    \and Max Planck Institute for Physics \label{mpi_add}
    \and INFN and Sapienza University of Rome \label{rom_add}
}

\date{Received: date / Accepted: date}
 
\maketitle


\begin{abstract}
A configurable calorimeter simulation for AI \\ (COCOA) applications is presented, based on the \textsc{Geant4} toolkit and interfaced with the \textsc{Pythia} event generator. This open-source project is aimed to support the development of machine learning algorithms in high energy physics that rely on realistic particle shower descriptions, such as reconstruction, fast simulation, and low-level analysis. Specifications such as the granularity and material of its nearly hermetic geometry are user-configurable. The tool is supplemented with simple event processing including topological clustering, jet algorithms, and a nearest-neighbors graph construction. Formatting is also provided to visualise events using the Phoenix event display software.


\end{abstract}

\section{Introduction}\label{sec:introduction}

Algorithms incorporating machine learning (ML) methods are a new paradigm in reconstruction, calibration, identification and analysis of High Energy Physics (HEP) experimental data. In recent years, various ML architectures have been deployed to optimize low-level tasks such as clustering, reconstruction, fast simulation, pileup suppression and object identification \cite{Guest:2018yhq,hepmllivingreview}. For example, ML-based fast calorimeter simulation relies on accurate target data to train a fast conditional generative model $p(D|T)$, where $T$ denotes the true set of stable final state particles produced in the collision and $D$ is the set of resulting detector hits. In particle reconstruction, on the other hand, the inverse process $D \rightarrow T$ is modelled by predicting a set of particles $R(D)$ to approximate $T$ as accurately as possible. The development of such algorithms requires a realistic, highly-granular simulation of particle detector response going beyond parameterized detector models frequently used in studies of particle physics phenomenology such as DELPHES~\cite{Mertens:2015kba}. 
In particular due to the complexity of particle showers in calorimeters, a detailed, microscopic simulation of interactions between particles and detector material is needed in order to develop low-level ML algorithms exploiting such features.

Recent research efforts to study calorimeter shower properties using ML~\cite{PhysRevD.97.014021,Qasim_2019,Di_Bello_2021} made use of the \textsc{Geant4}~\cite{ALLISON2016186} simulation toolkit for simple detector geometries. However an open source simulation with a realistic cylinder like hermetic geometry and realistic detector is yet to be adapted by HEP community for ML studies and beyond.
Aiming to bridge this gap, the COnfigurable Calorimeter simulatiOn for Ai (\scd{}) was developed, which uses \textsc{Geant4}~\cite{ALLISON2016186} to implement detailed shower simulation for particles in a full-coverage, highly-segmented sensitive volume comparable to that of multipurpose detectors at the LHC. 
The program source code \cite{sanmay_ganguly_2023_7700475} is linked together with a technical documentation on the project website\footnote{\url{https://cocoa-hep.readthedocs.io/en/latest/index.html}}. 
The emphasis of this software package is on realistic calorimeter simulation. No realistic digitization and electronic readouts are implemented and energy loss due to these processes are neglected in this package. For the same reason, simplified tracking is included in \scd{} to model particle deflection in a magnetic field and energy depositions upstream of the calorimeter. A sophisticated open model for tracking suitable for tracking studies based on silicon hits is provided by \cite{opendatadetector}.

\begin{figure}[!t]
    \centering
    \includegraphics[width=0.50\textwidth]{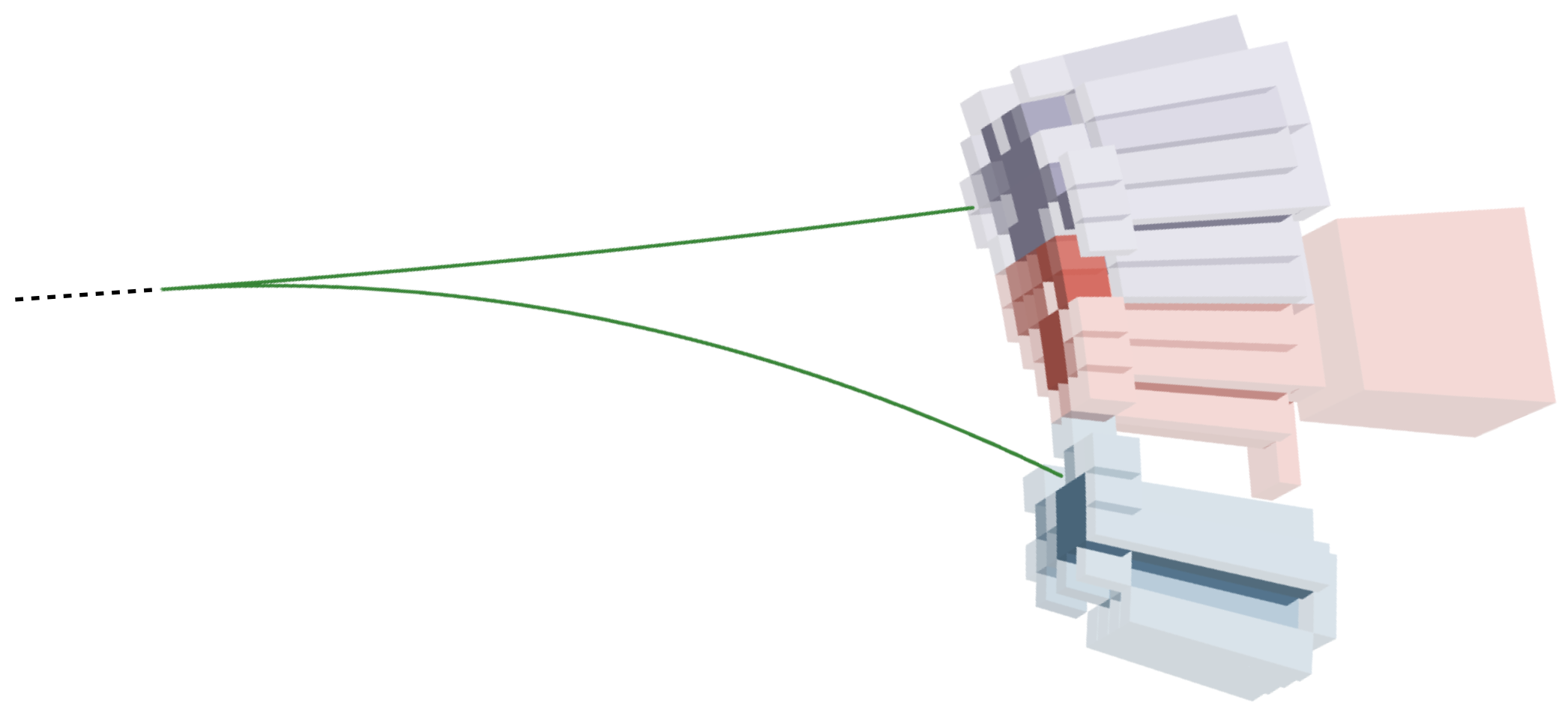}
    \caption{Visualization of a photon (dashed line) with energy 50 GeV converting to two electrons (green lines) producing three distinct clusters in the \scd{} central electromagnetic calorimeter. The cluster shown in red contains an additional cell in the first layer of the hadronic calorimeter due to a noise fluctuation. Cells are shown with an opacity proportional to energy over noise ratio divided by 4.6, the threshold for topoclustering seeds.}
    \label{fig:photonconversion}
\end{figure}

Usability for ML-based studies is a core motivation in the design of the \scd{} code.
Datasets generated by \scd{} have featured in two recent applications of ML to 
particle reconstruction and fast simulation \cite{DiBello:2022rss,DiBello:2022iwf}.
To this end, the main parameters of the calorimeters are largely configurable, including their material, granularity, depth and the amount of readout noise. 
Similarly the inclusion of material interactions in the tracking region is optional.
For comparisons with benchmark reconstruction approaches, output data from \scd{} are conveniently interfaced to standard topological clustering and jet clustering algorithms. 
The output includes a record of energy contributions to each cell by truth particles for supervising cell-level predictions and edge lists for connecting cells and tracks in a graph to support geometric deep learning models.
Finally, the default geometry has been formatted for rendering in the Phoenix event display software \cite{phoenix}, along with a script to export event output files for visualization. An example is shown in Fig. \ref{fig:photonconversion}.
%
%

The sophisticated \scd{} calorimeter simulation and its data post-processing provides users easy access
to datasets suitable to train models for current collider experiments
or for more general algorithms development and benchmarking.
On the other hand, the open-source nature of the package and its visualization support have the potential
for use cases in education and science communication in HEP.

\section{Detector design}\label{sec:detectorDesign}
\begin{figure*}[ht!]
    \centering
    \subcaptionbox{\label{fig:SCD-YZ}}[0.55\textwidth]{\includegraphics[width=1.0\linewidth]{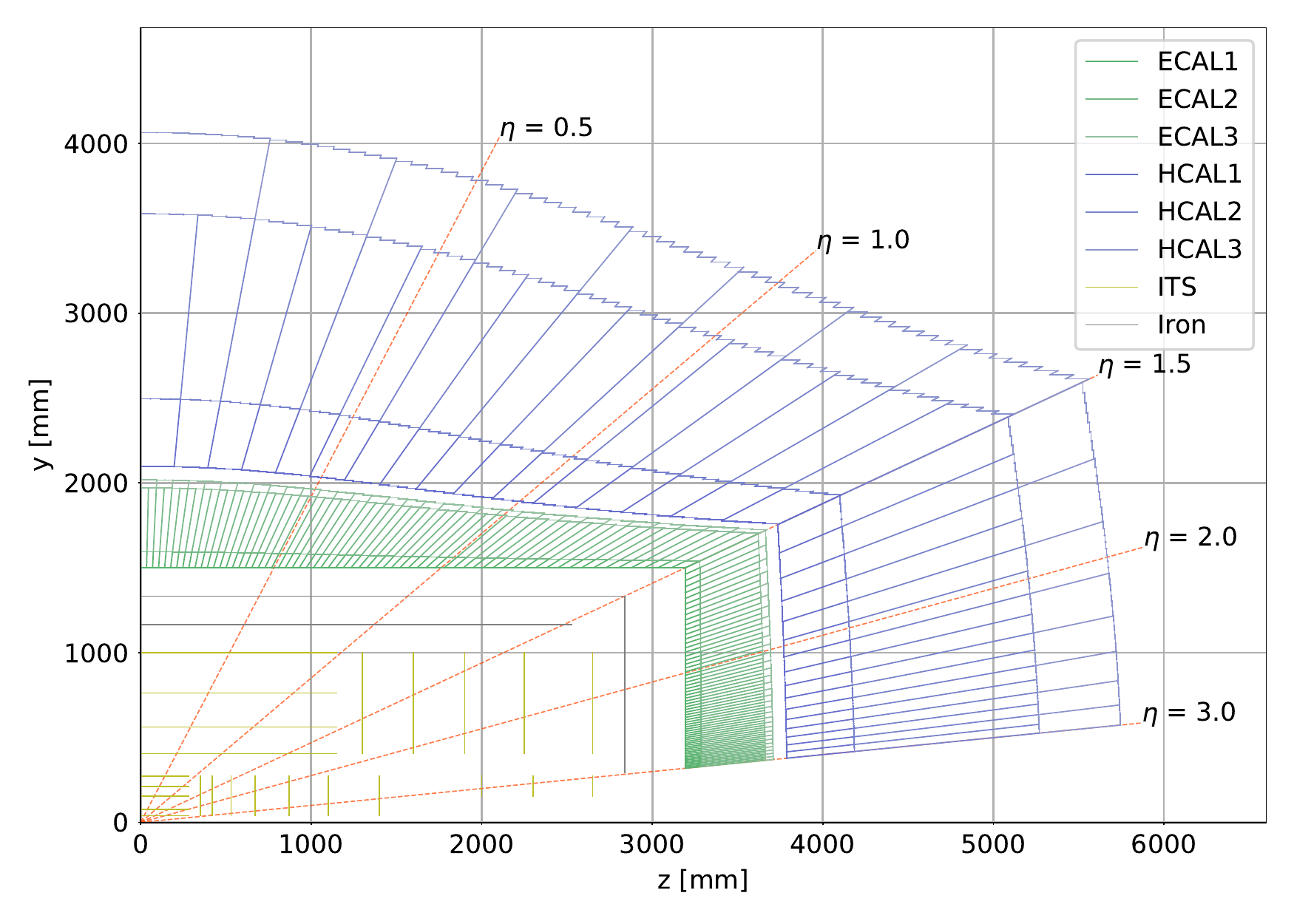}}
    \quad
    \subcaptionbox{\label{fig:SCD-XY}}[0.385\textwidth]{\includegraphics[width=1.0\linewidth]{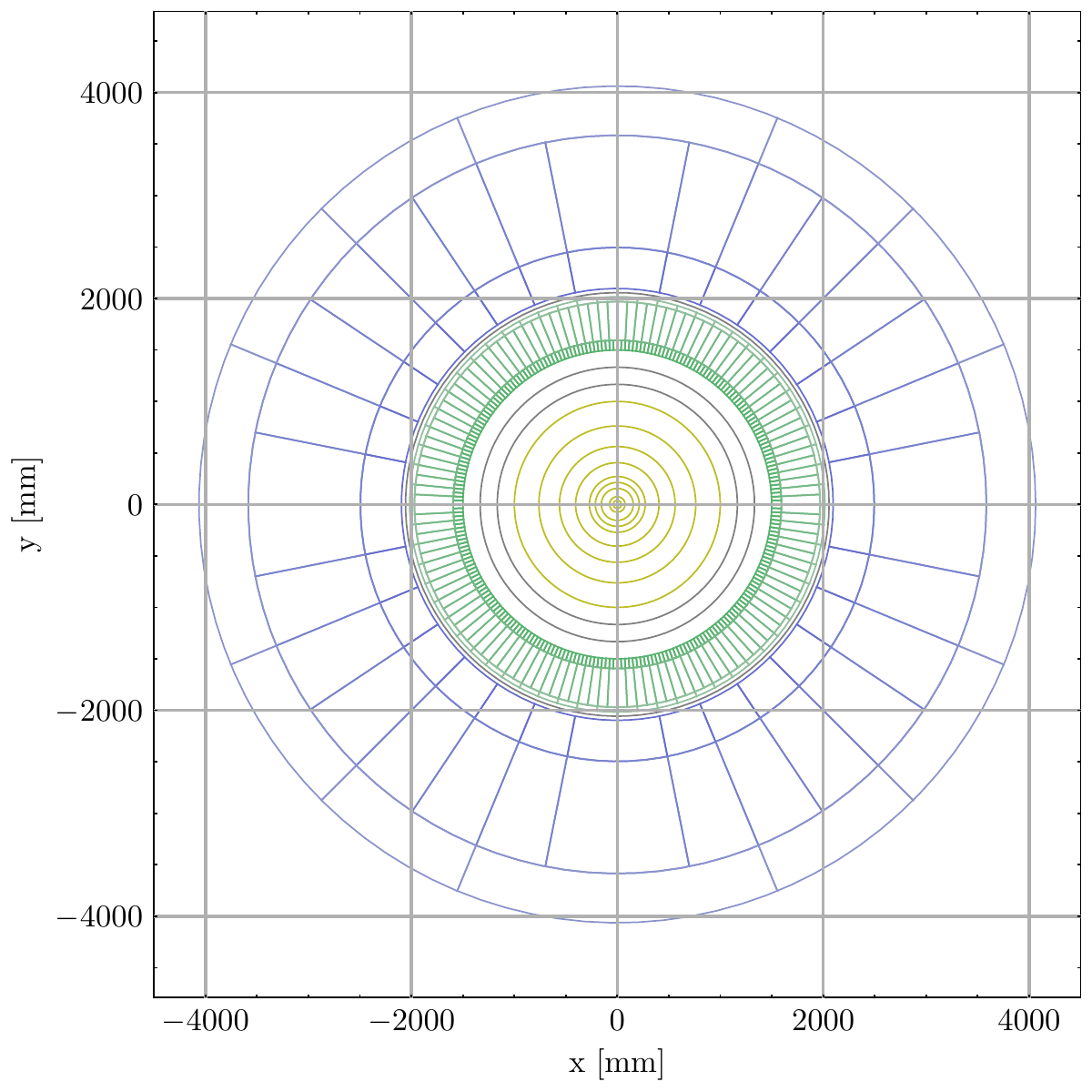}}
    \caption{Positive quadrant scheme of \scd{}. We use a right-handed orthogonal coordinate system $x$-$y$-$z$, where $z$-axis is the principal axis of the detector and a constant $z$ refers to a circular cross-section of the detector. (a) $yz$-projection showing the \scd{} ITS, subsequent iron layers, calorimeter system in the barrel and end-cap region, overlaid on lines marking constant pseudorapidity $\eta$. (b) $xy$-projection shows barrel region of the same subsystems at $z=0$.}
    \label{fig:Barrel_EndCap_scheme}
\end{figure*}

\begin{figure*}[ht!]
    \centering
    \includegraphics[width=1\textwidth]{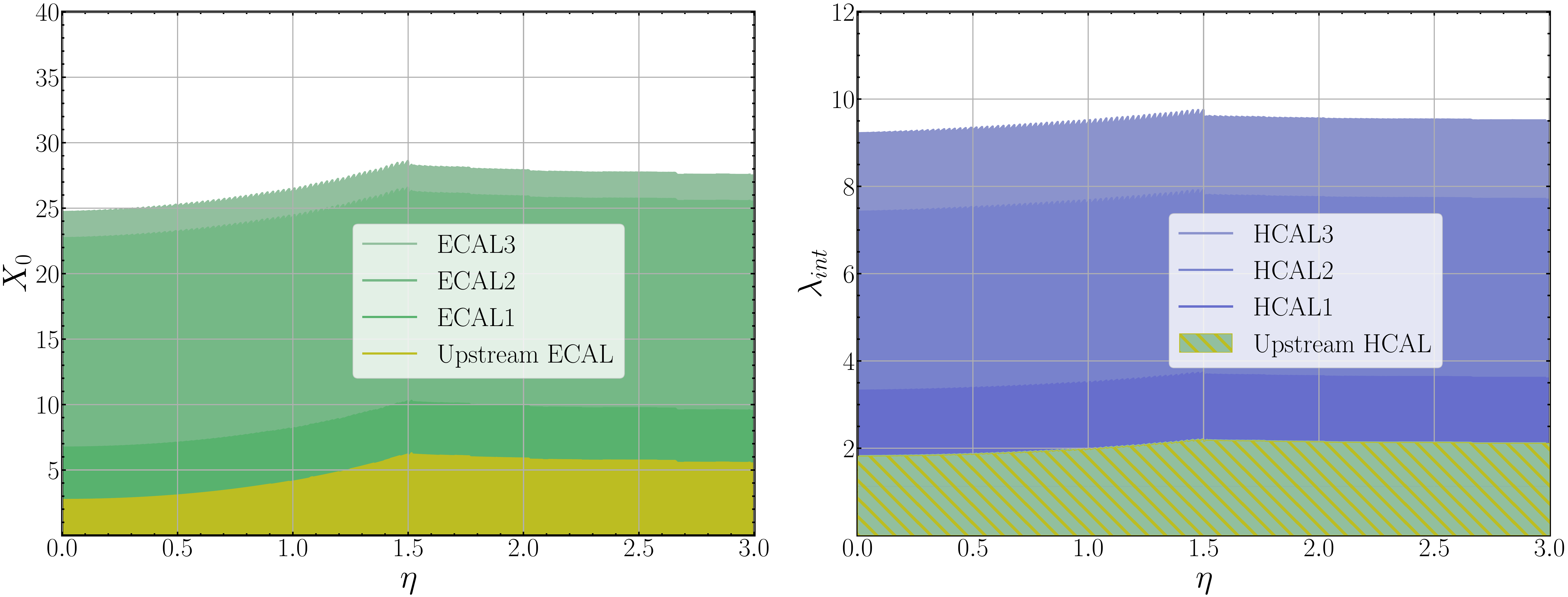}
    \caption{(left) Cumulative amounts of material, in units of radiation length $X_0$ and as a function
of $\eta$, in front of and within the electromagnetic calorimeter system. (right) Cumulative amounts of material, in units of interaction length $\lambda_{int}$ and as a function
of $\eta$, in front of and within the hadronic calorimeter system.
}
    \label{fig:Rad_Int_length}
\end{figure*}

The major components of \scd{} are an inner tracking system (ITS) surrounded by an electromagnetic calorimeter (ECAL) and finally a hadronic calorimeter (HCAL). These subsystems are arranged concentrically and are symmetric in azimuthal angle $-\pi < \phi \leq \pi$ as shown in Fig.~\ref{fig:Barrel_EndCap_scheme}. 
No muon spectrometer is considered in this design and muons are reconstructed as tracks with the ITS. The goal of this design is to accurately model the relevant outputs of a multipurpose detector at the LHC while being simplified by the exclusion of detailed components like readout electronics, cabling, and support structures. The detector design is largely configurable, with its default parameter values chosen to achieve response characteristics comparable to that of the ATLAS detector. Following is a detailed description of each subsystem.

The ITS consists of hollow cylinders in the central detector part and disks at both of its ends, each of which are centered around the beamline.
Each of these components consists of a silicon layer of 150~\mum thickness in case of the disks and the five innermost cylinders and 320~\mum in case of the 4 outermost cylinders.
Each silicon layer is accompanied by an iron layer of 350~\mum thickness in order to provide a simulate support material. The ITS only serves the purpose of simulating the interaction of particles with matter upstream of the calorimeter. The resulting detector hits
are not used for tracking purposes.
The default value of the magnetic flux density present in the ITS amounts to 3.8~T.
Finally, two layers of iron totalling 4.4 cm in depth are added to represent support or cryostat material in front of the calorimeter.

The inner surface of the calorimeter system is a cylinder with a radius of 150 cm and a length of 6387.8 cm immediately enclosing the iron layers and the ITS. The calorimeters are separated into a central barrel region covering the pseudorapidity range $|\eta|<1.5$ and two end-cap regions extending the coverage up to $\eta=3$ by default. 
Both the ECAL and the HCAL are divided into 3 concentric layers, with each layer being further segmented into cells with edges of constant $\eta$ and $\phi$.
The cell granularity for each layer is configurable by setting the number of equal divisions in $\eta$ and (separately) $\phi$. 
The depth of the cells in every layer is designed to be nearly constant in $\eta$ to ensure that the fraction of a particle's energy deposited in each layer does not depend on the incident angle. This design, leading to layer shapes of the form $1\,/\,\cosh{\eta}$, provides a uniform calorimeter thickness as a function of pseudo-rapidity. \scd{} will thus have a more uniform response than a pure circular cylindrical shape. 

The \scd{} calorimeter material is a compound using an equivalent molecule approximation, mixing an absorber and scintillator material with a constant proportion.
Both the materials and their proportion can be configured for the ECAL and the HCAL individually. By default, the ECAL is made of a mixture of lead and liquid argon, corresponding to the ATLAS ECAL materials.
The volume proportion amounts to $1\!:\!3.83$, resulting in a radiation length of $X_0 = 2.5$ cm.
The ECAL and HCAL are separated by an iron layer with a default thickness of 80\,mm. The HCAL is made of a mixture of iron and polyvinyl toluene plastic material with a volume proportion of  $1.1.:1.0$, resulting in a nuclear interaction length of $\lambda_{int}=26.6$ cm.
The integrated radiation and interaction length measured from the IP to the end of the HCAL is shown as a function of the pseudorapidity in Fig~\ref{fig:Rad_Int_length}.

\begin{table}[]
    \centering
    \caption{
    Calorimeter default design values regarding layer depths in terms of radiation lengths $X_0$ (ECAL) and hadronic interaction lengths $\lambda_{int}$ (HCAL), granularity and energy noise levels.
    }
    \begin{tabular}{cccc}
    \toprule
        Layer    & Depth &  \begin{tabular}{@{}c@{}}Segmentation \\($\eta \times \phi$)\end{tabular} & \begin{tabular}{@{}c@{}}Std. dev. noise per cell \\per event {[}MeV{]}\end{tabular} \\
        \midrule
        ECAL 1 & \hspace{1.2ex}$4\,X_0$   & $256 \times 256$ & 13               \\
        ECAL 2 &               $16\,X_0$  & $256 \times 256$ & 34               \\
        ECAL 3 & \hspace{1.2ex}$2\,X_0$   & $128 \times 128$ & 41 \\
        \midrule
        HCAL 1 & $1.5\,\lambda_{\mathrm{int}}$ & $64 \times 64$ & 75               \\
        HCAL 2 & $4.1\,\lambda_{\mathrm{int}}$ & $64 \times 64$ & 50               \\
        HCAL 3 & $1.8\,\lambda_{\mathrm{int}}$ & $32 \times 32$ & 25               \\
        \bottomrule
    \end{tabular}
    \label{tab:CaloLayers}
\end{table}

While this calorimeter design represents a homogeneous detector, a spread in the resolution of reconstructed energies in accordance with a sampling calorimeter design is emulated by means of configurable sampling fraction parameters for the ECAL and the HCAL individually. In lieu of a complete simulation of active and passive material, the sampling is emulated by accounting only for a fraction of the GEANT energy deposits steps for all particles in the calroimeter showers. The steps to removed are chosen randomly. The sum of the total deposited energy by those steps is computed and  the total energy released is estimated by inverse scaling of the total deposited energy by the corresponding fraction.

Noise, as for example from electronics, is simulated by the addition of random amounts of energy following a Gaussian distribution centered around zero. The noise is independently added to each cell. Negative energies are allowed as is typically the case as a result from the subtraction of pedestals. If such downward fluctuations are significant in size, those negative energy cells can be clustered into topoclusters. 
The default choices of materials and smearing parameters provided in Tab.~\ref{tab:CaloLayers} are chosen in order to approximate single-particle responses of the ATLAS calorimeter system~\cite{ATLAS:2021epk}.

\section{Data processing}\label{sec:dataProcessing}

\begin{figure}[!t]
    \centering
    \includegraphics[width=0.49\textwidth]{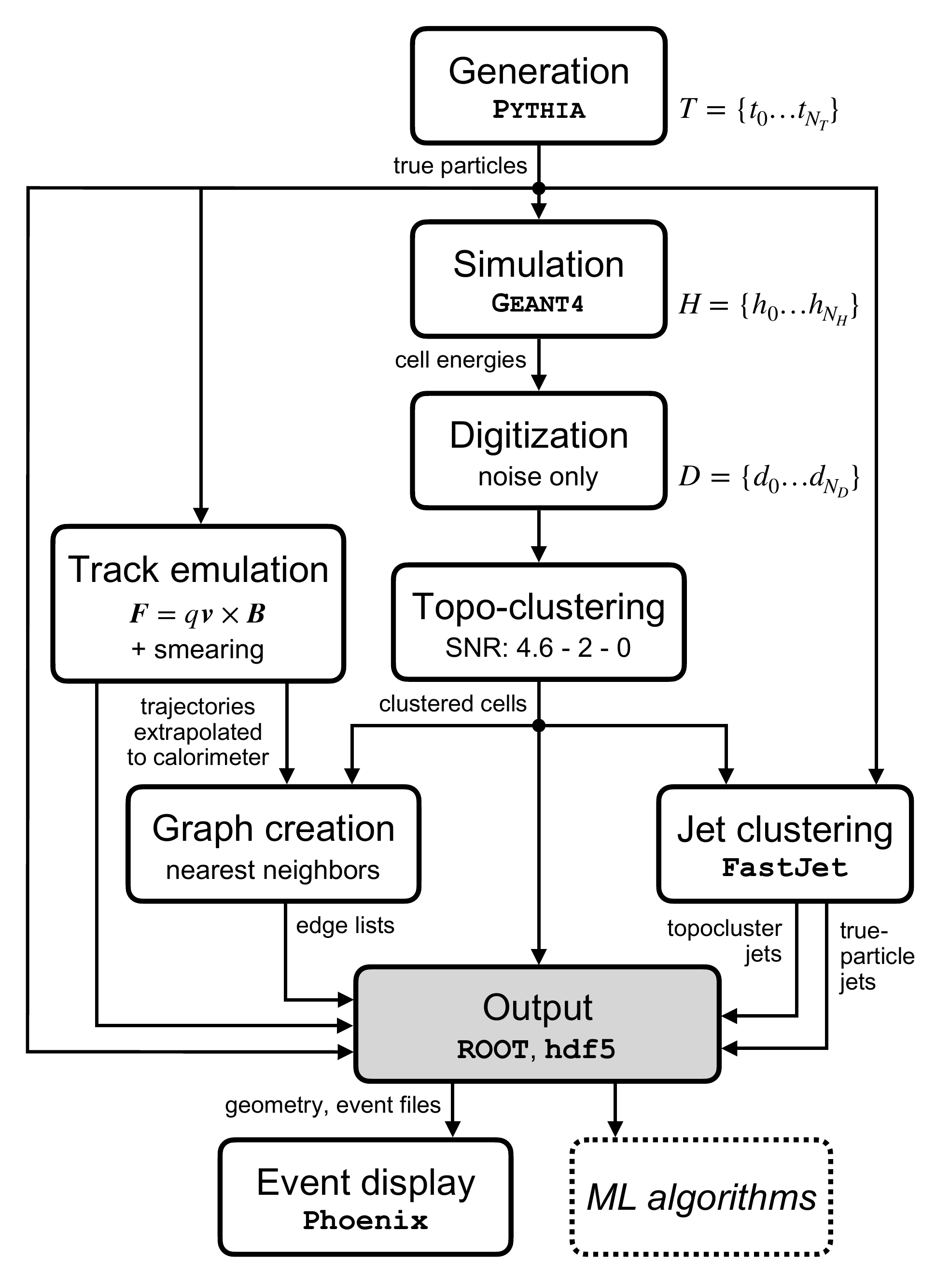}
    \caption{\scd{} workflow. Primary particles generated with the \textsc{Pythia} library are introduced to \scd{}.
    Their interactions with the detector material is simulated by means of the \textsc{Geant4} toolkit.
    Calorimeter cells identified by a topological clustering algorithm are stored in the output \texttt{ROOT} file together with true particles, emulated tracks, and particle trajectories extrapolated from the IP through the calorimeter according to the equations of motion. A nearest-neighbors-based graph is constructed and stored via edge lists connecting source and destination nodes amongst the output cells and tracks. Jets made of true particles as well as topoclusters are stored in the output file as well. Events in the output file can be parsed for visualization in \textsc{Phoenix}.}
    \label{fig:workflow}
\end{figure}


\begin{figure*}[!h]
    \centering
    \subcaptionbox{\label{fig:TopoN}}[0.45\textwidth]{\includegraphics[width=1.0\linewidth]{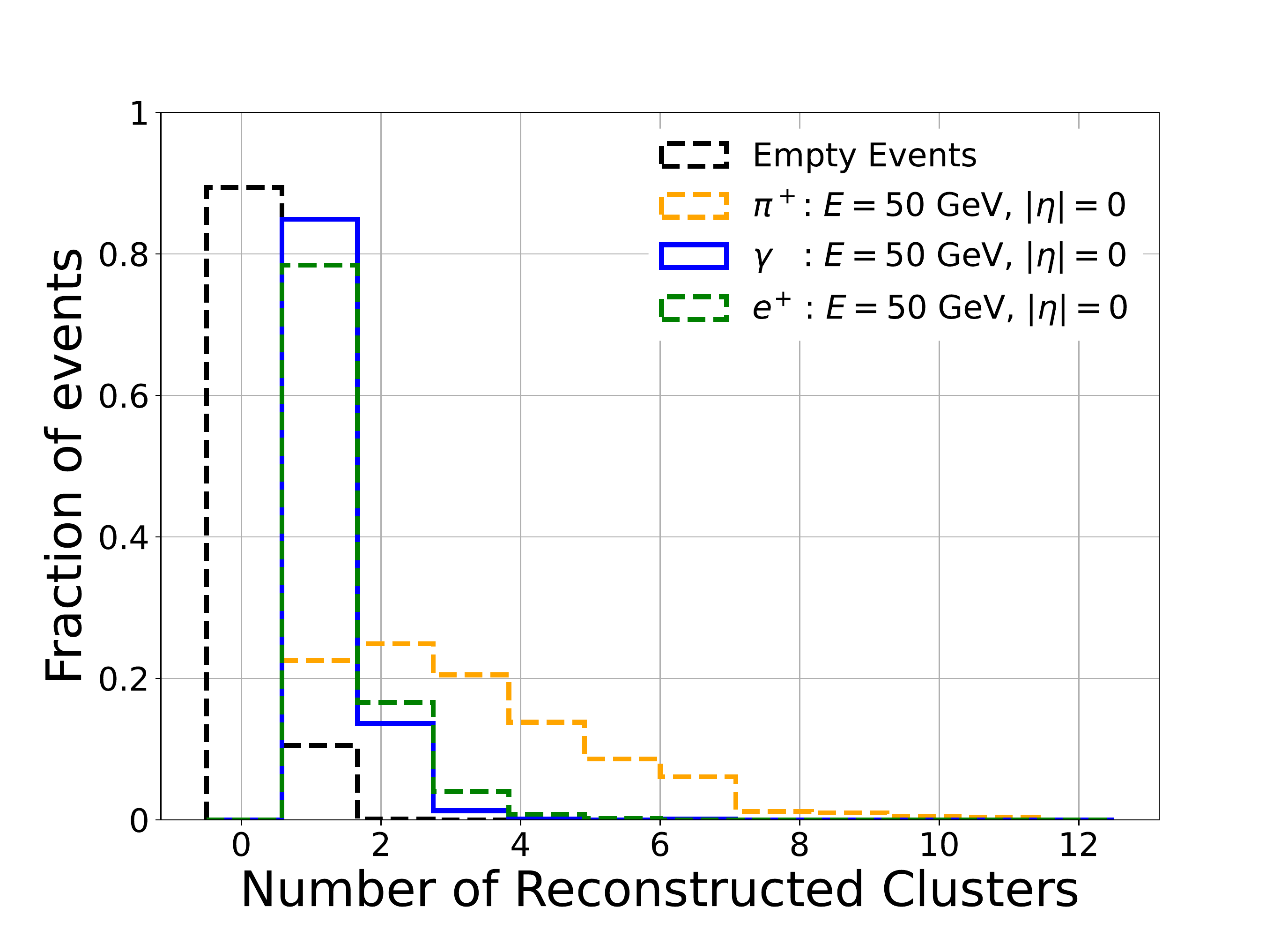}}
    \hfill
    \subcaptionbox{\label{fig:TopoE}}[0.48\textwidth]{\includegraphics[width=1.0\linewidth]{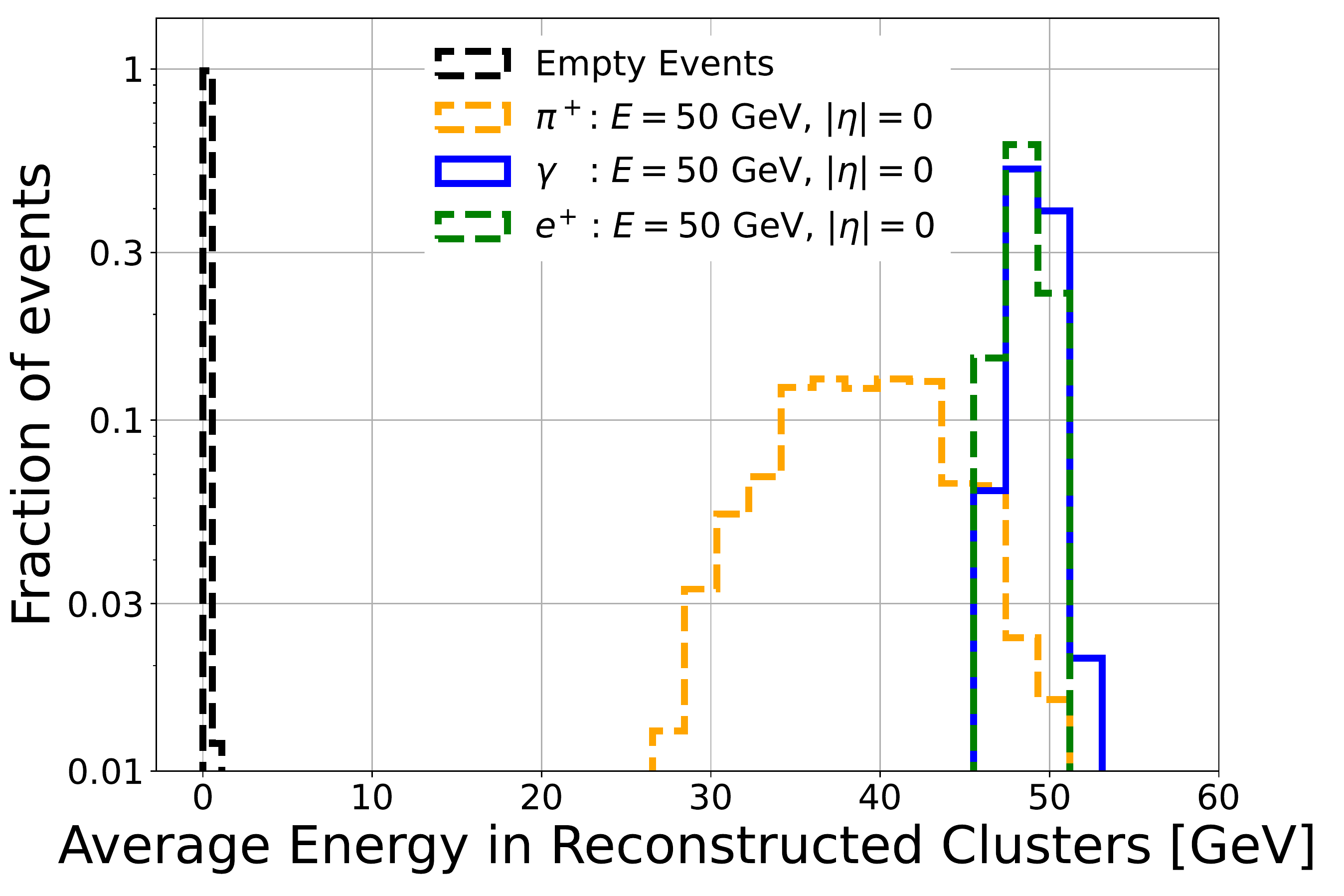}}
    \caption{Number (a) and average energy (b) of reconstructed clusters in \scd{} for events with a single charged pion, electron, or photon shot at $\eta = 0$. Results are also shown for clusters reconstructed in empty events due to electronic noise. The default topological calorimeter cell clustering settings are used.}
    \label{fig:Topo}
\end{figure*}

\noindent
Every event is processed according to the workflow presented in Fig.~\ref{fig:workflow}.
First, primary particles are generated at the IP by means of the \textsc{Pythia8} Monte Carlo event generator~\cite{pythia}.
A broad range of primary physics processes is available to the user, ranging from the generation of single particles as well as single jets up to more complicated final states with large multiplicities of jets and leptons in the final state.

The set of final state, stable particles is stored in the output file and passed on to the detector simulation
described in the previous section, where the propagation of these particles and their interactions with the detector material is simulated in \textsc{Geant4}~\cite{ALLISON2016186}.
The model of hadronic interactions is chosen in accordance with the ATLAS and CMS detector simulations.
The sum of the energies deposited in each calorimeter cell is stored. 
Electronic noise is simulated by the addition of random energy offsets to each cell for which Tab.~\ref{tab:CaloLayers} provides the default values of standard deviation for each layer.

For the purpose of particle reconstruction, the origin of energy deposits in each cell is stored via a list of parent particle indices which contributed energy into the cell and a list of weights recording their relative contribution to the total cell energy.
Cells which received their dominant contribution from electronic noise are assigned an index of -1.

In order to limit the number of calorimeter cells stored in the output file to a reasonable level, low-energy cells dominated by noise contributions are suppressed using a topological clustering algorithm \cite{Aad_2017}. 
Only cells contained in the resulting ``topoclusters'' are stored in the output file.
Topoclusters are seeded by single cells which are required to contain a deposited energy well above the noise level, where the threshold of this signal-to-noise ratio (SNR) is  4.6 for \scd{} by default, while a value of 4.0 is used for the ATLAS experiment.
This difference is chosen in order to achieve a better agreement between ATLAS and \scd{} in terms of the topocluster multiplicity distribution for single charged and neutral pions as well as pure noise events (Fig.~\ref{fig:Topo}).
Starting with the seeding cells, all neighbouring cells are added to the cluster if their SNR is above another threshold, where the default value is set to 2.
Finally, all further neighbouring cells above a third threshold are added, which by default is set to 0.
Cells with negative energy can be included, based on their absolute value, or excluded entirely (default configuration).
Topocluster candidates containing multiple local maxima in ECAL cell energy each surpassing 400 MeV are split into separate topoclusters. 

In order to support particle reconstruction studies which include high energy primary photons, electron-positron pairs from photon conversions taking place in the ITS upstream its two outermost iron layers are stored in the \scd{} output file as well. Tracks emanating from photon conversions and also primary electron tracks are used to construct groups of topoclusters denoted ``superclusters'' associated with electron and photon showers. The superclustering procedure in \scd{} follows the criteria described in \cite{Aad2019-vd}, designed to improve electron energy reconstruction by incorporating nearby energy deposits from bremsstrahlung. It also includes criteria for grouping multiple clusters that are related by a pair of nearby tracks to a photon conversion vertex, thus improving reconstructed photon energy. In the photon conversion shown in Fig. \ref{fig:photonconversion}, for example, the \scd{} output contains a supercluster which combines the three topoclusters shown. Due to the simplified tracking, the criteria on number of track hits are not applied. The \scd{} implementation does not focus on electron and photon identification; rather, superclusters are only formed using tracks linked to primary or conversion electrons.

\begin{table}[]
    \centering
    \caption{
    Default $k$ (number of nearest-neighbors) and maximum $\Delta R$ separation used to define edges in the fixed graph creation. Edges between cells are denoted ``c-c'' while edges between tracks and cells are denoted ``t-c''.
    }
    \begin{tabular}{c|ccc|ccc}
        \multicolumn{1}{c}{} & \multicolumn{3}{c}{ECAL layer} & \multicolumn{3}{c}{HCAL layer} \\
        \cmidrule{2-7}
        \multicolumn{1}{c}{} & 1 & 2 & 3 & 1 & 2 & 3 \\
        \toprule
        $k$ (c-c) inter-layer & 1 & 2 & 2 & 2 & 2 & 1 \\
        $k$ (c-c) same layer & 8 & 8 & 8 & 6 & 6 & 6 \\
        $k$ (t-c) & 4 & 4 & 4 & 3 & 3 & 3 \\
        \midrule
        $\Delta R^{\mathrm{max}}$ (c-c) & 0.05 & 0.07 & 0.14 & 0.30 & 0.30 & 0.60 \\
        $\Delta R^{\mathrm{max}}$ (t-c) & 0.15 & 0.15 & 0.40 & 1.10 & 1.10 & 2.00 \\
        \bottomrule
    \end{tabular}
    \label{tab:GraphCreation}
\end{table}

While the event simulation based on the \textsc{Geant4}~\cite{ALLISON2016186} toolkit determines particle trajectories according to their equations of motion and their interactions with detector material, \scd{} implements a particle tracking based only on the equations of motion for the benefit of downstream tasks. These tracks are extrapolated to the entry surface as well as each layer of the calorimeter and the resulting $\eta$ and $\phi$ coordinates are stored in the output file. 

For user convenience, an interface to the FastJet~\cite{Cacciari_2012} library is provided that clusters primary particles as well as topological calorimeter cell clusters into jets. 
The user can choose the specific jet clustering algorithm accordingly, with the anti-$k_\mathrm{T}$ algorithm set as default.

For each event in the output data a fixed heterogeneous graph containing cells and tracks is provided by means of two lists storing the indices of source and destination nodes for each edge.
The edges are created based on $k$ nearest neighbors in angular distance with $k$ being user-configurable per calorimeter layer and edge type. 
Three edge types are defined: track-to-cell, cell-to-cell inter-layer, and cell-to-cell across neighboring calorimeter layers (tracks are not directly connected).
The user can configure for each of these types both how many edges to construct in a $\Delta R$-ordered neighborhood and also with a maximum $\Delta R$ (where $\Delta R^2 = \Delta \eta^2 + \Delta \phi^2$). The default values are given in Tab. \ref{tab:GraphCreation}.

The final output file produced by \scd{} stores an array of features for each event which are associated with the following sets: cells that participated in topoclusters, tracks, topoclusters, truth particles and decay record, graph edges, and jets. The output file format is \texttt{ROOT} but can be converted to \texttt{hdf5} format using a script provided in the repository.

\begin{figure}
    \centering
    \includegraphics[width=0.46\textwidth]{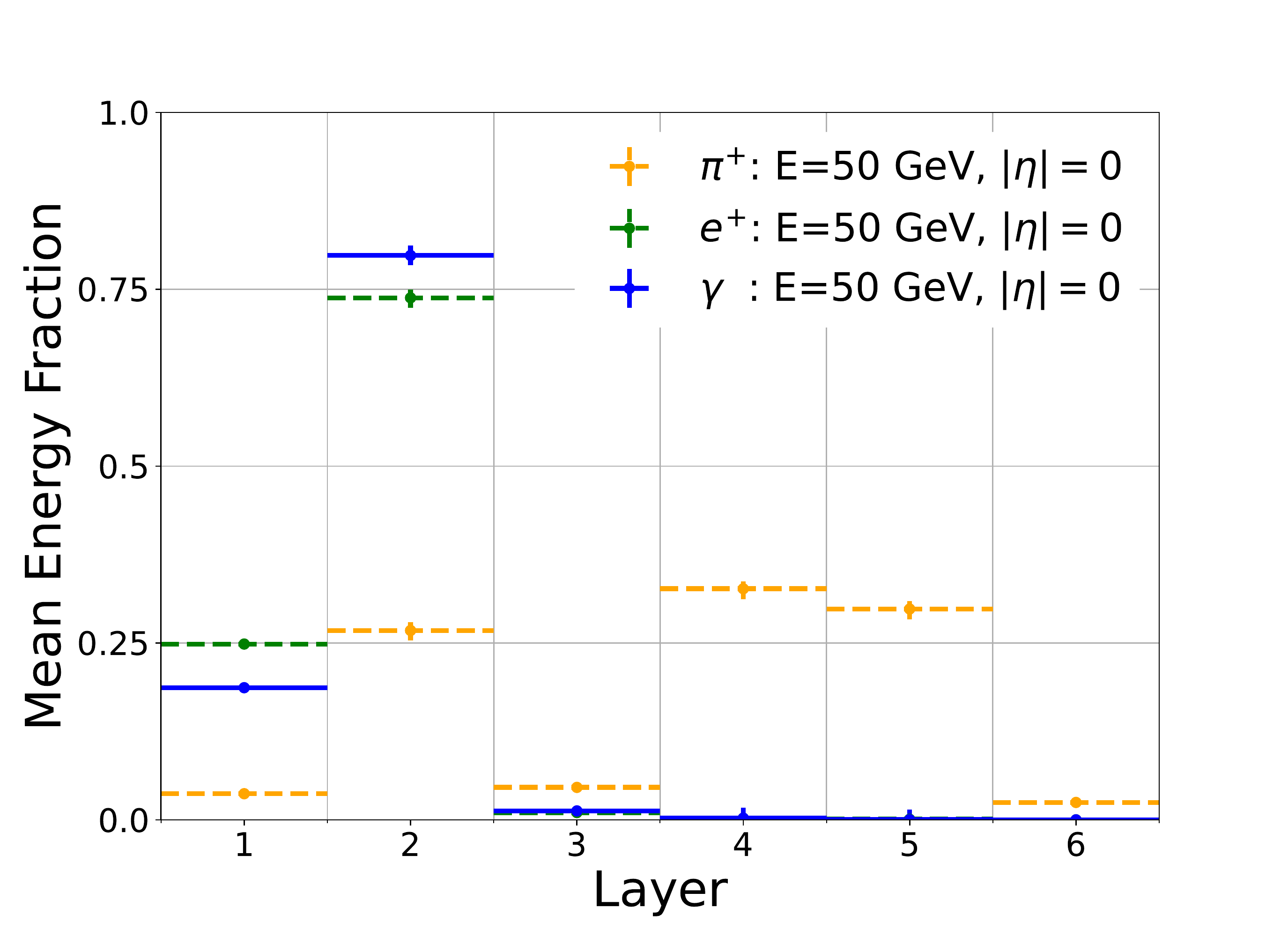}
    
    \caption{Energy deposited by electrons, photons and charged pions
    for each calorimeter layer. The electron and photon showers are limited
    to the electromagnetic calorimeter (layers 1 to 3) while the charged pion showers
    reach deep into the hadronic calorimeter (layers 4 to 6)}.
    \label{fig:layerFractions}
\end{figure}

\begin{figure*}[btp!]
    \centering
    \subcaptionbox{\label{fig:eResElectrons}}[0.33\textwidth]{\includegraphics[width=
    \linewidth]{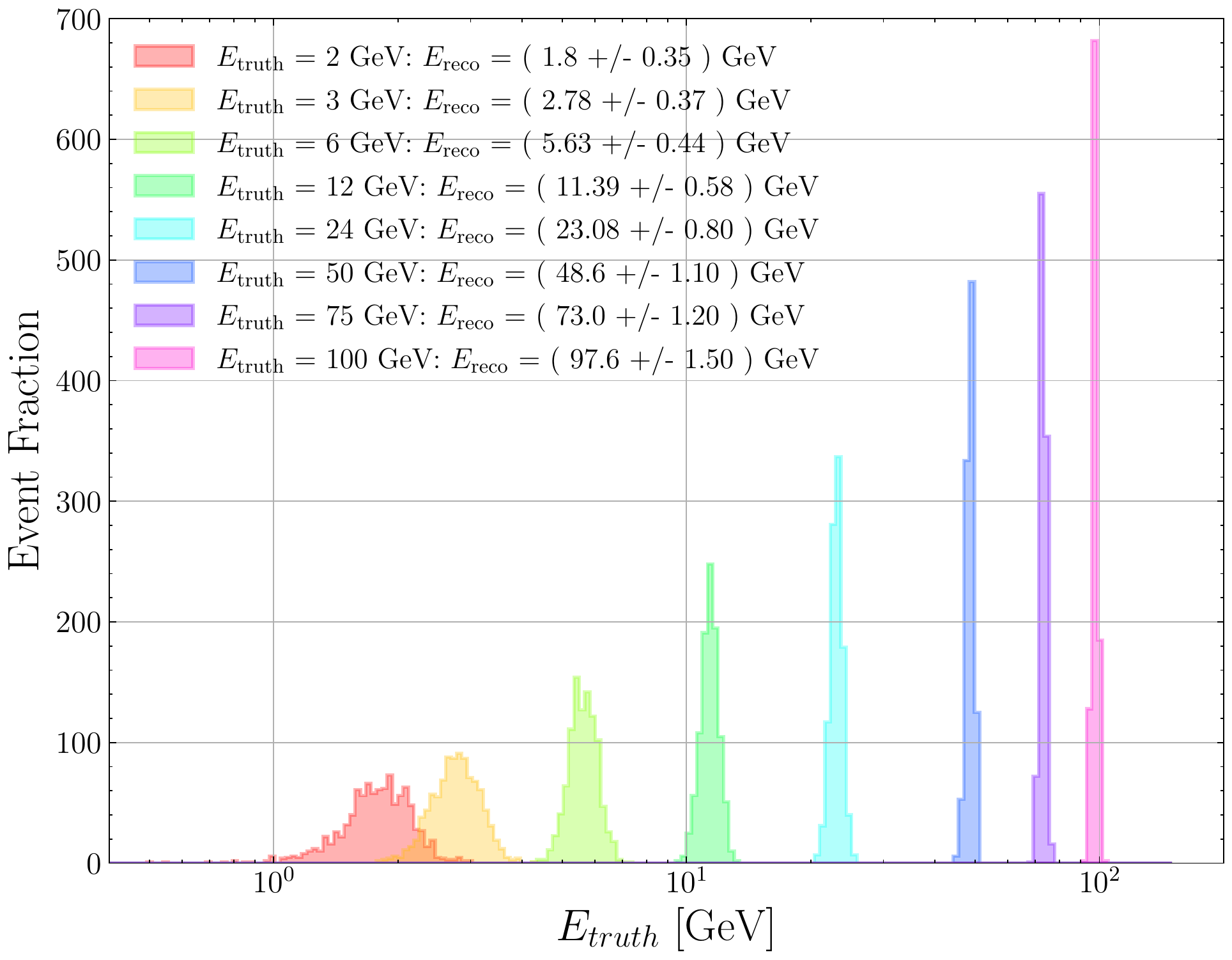}}
    \subcaptionbox{\label{fig:eResPhotons}}[0.33\textwidth]{
    \includegraphics[width=\linewidth]{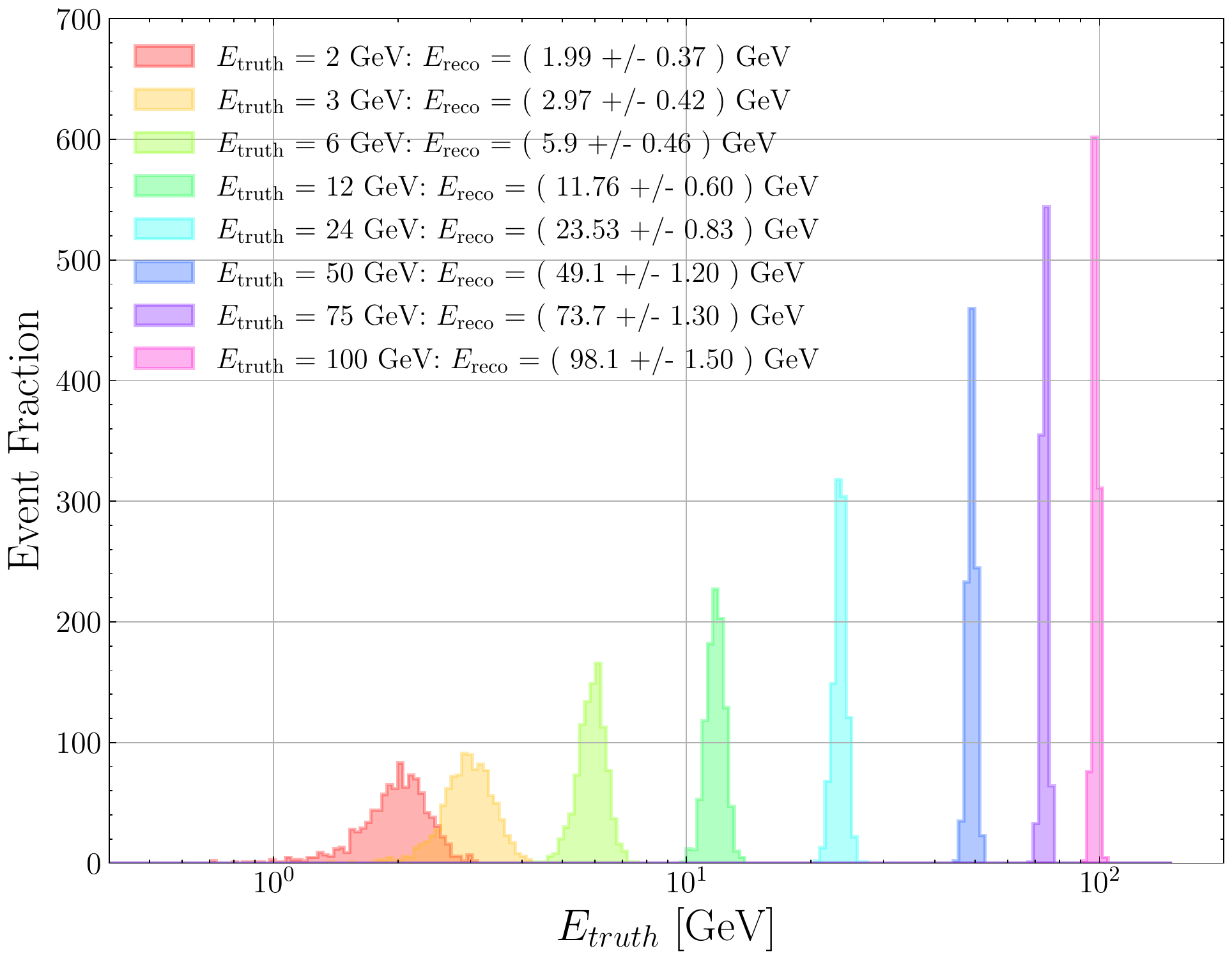}}
    \subcaptionbox{\label{fig:eResChargedPions}}[0.33\textwidth]{
    \includegraphics[width=\linewidth]{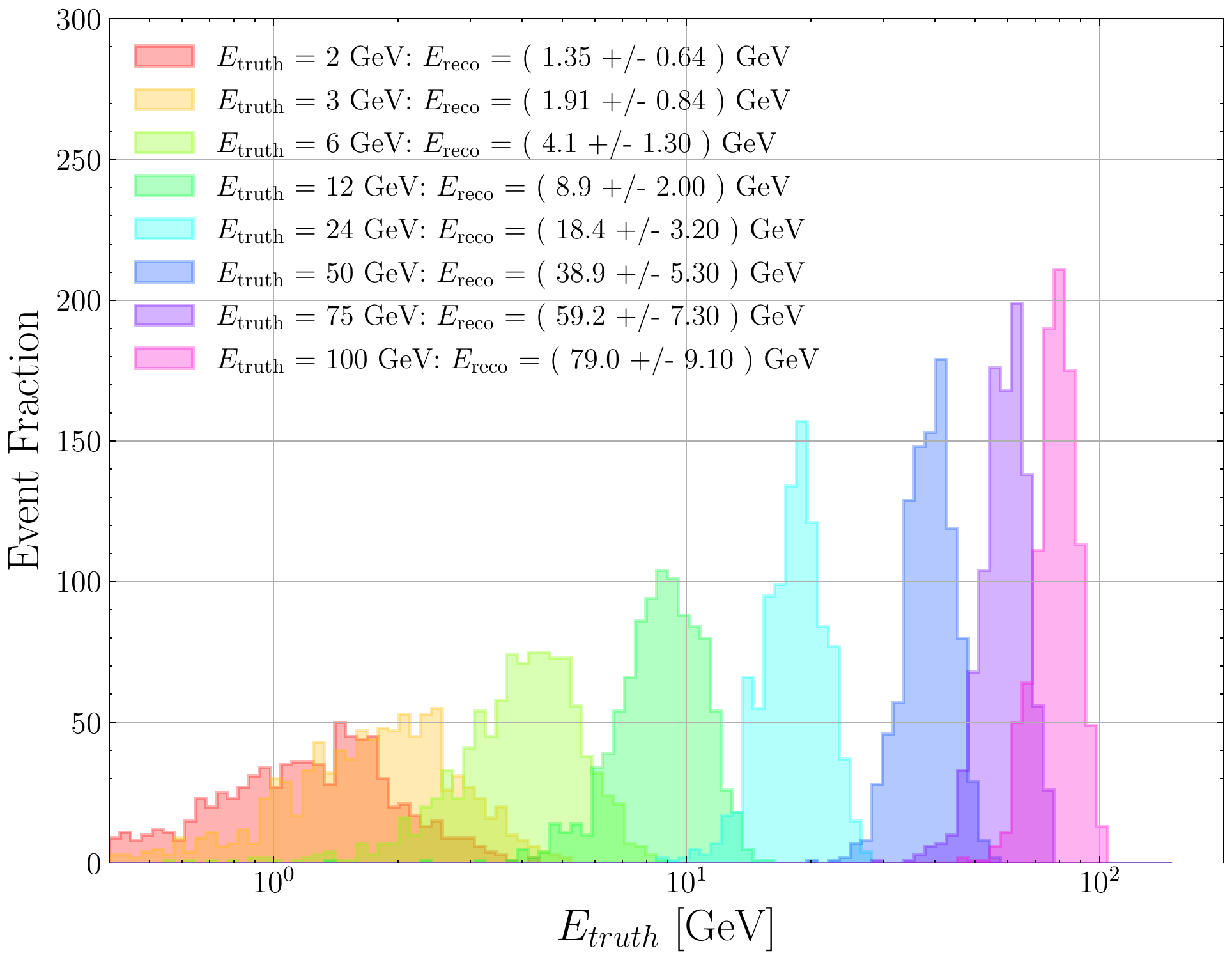}}
    \caption{Reconstruction energies for central electrons (\ref{fig:eResElectrons}), photons (\ref{fig:eResPhotons}) and
    charged pions (\ref{fig:eResChargedPions}). The average reconstructed energies
    follow the initial particle energies while the energy resolution improves as the initial particle energy increases.   }
    \label{fig:eResponses}
\end{figure*}

\section{Detector performance}\label{sec:performance}
In the following, the performance of \scd{} is investigated by means
of single particles which are generated at the IP.
For each particle type and momentum under investigation,
the event generation is repeated in order to gather
a statistically significant amount of events.

The correct reconstruction of particle energies is demonstrated in
Fig.~\ref{fig:Topo},
which compares the distributions of multiplicities (Fig.~\ref{fig:TopoN})
and energy sums (Fig.~\ref{fig:TopoE}) of topoclusters for charged pions, photons, electrons and events containing only noise contributions, denoted as empty events.
In most of the empty events, the cell energies do not pass
the noise threshold of the clustering algorithm. For those events in which this threshold is passed, the average cluster energy sum amounts to 36 MeV in line with the low noise levels provided in Tab.~\ref{tab:CaloLayers}.
The photons and electrons mostly result in one cluster, while their energy is reconstructed with only
a small variation. In comparison, the charged pion events
result in larger variations of the cluster multiplicity and energy sum distributions due to the higher degree of variations in deposited energies
for the hadronic showers.
The average cluster energy sum is below the initial charged pion energy due to the involved nuclear interactions of the shower particles with the detector material, which are not counted as detectable energy.
A hadronic calibration procedure is not performed within \scd{} but left for
downstream tasks.



Patterns of energy depositions across the calorimeter
are demonstrated in Fig.~\ref{fig:layerFractions} in terms of fractions of deposited energy per calorimeter layer for electrons, photons and charged pions.
As a consequence of the material budget presented above in
Fig.~\ref{fig:Rad_Int_length}, the electrons and photons deposit most of
their energy in the electromagnetic calorimeter, in particular in the second
calorimeter layer, while the charged pions reach the hadronic calorimeter layers
where they deposit most of their energy, in line with energy deposition patterns
at collider-detector experiments.

Figure~\ref{fig:eResponses} shows distributions of the reconstructed energies 
for central electrons, photons and charged pions with different initial energies.
The energy resolution provided by the calorimeter response
improves as the initial particle energy increases.
It is larger for charged pions compared to electrons and photons, 
as expected because of the existence of large sampling fluctuations 
for hadronic showers compared to electromagnetic showers.

\begin{figure}
    \centering
    \includegraphics[width=0.46\textwidth]{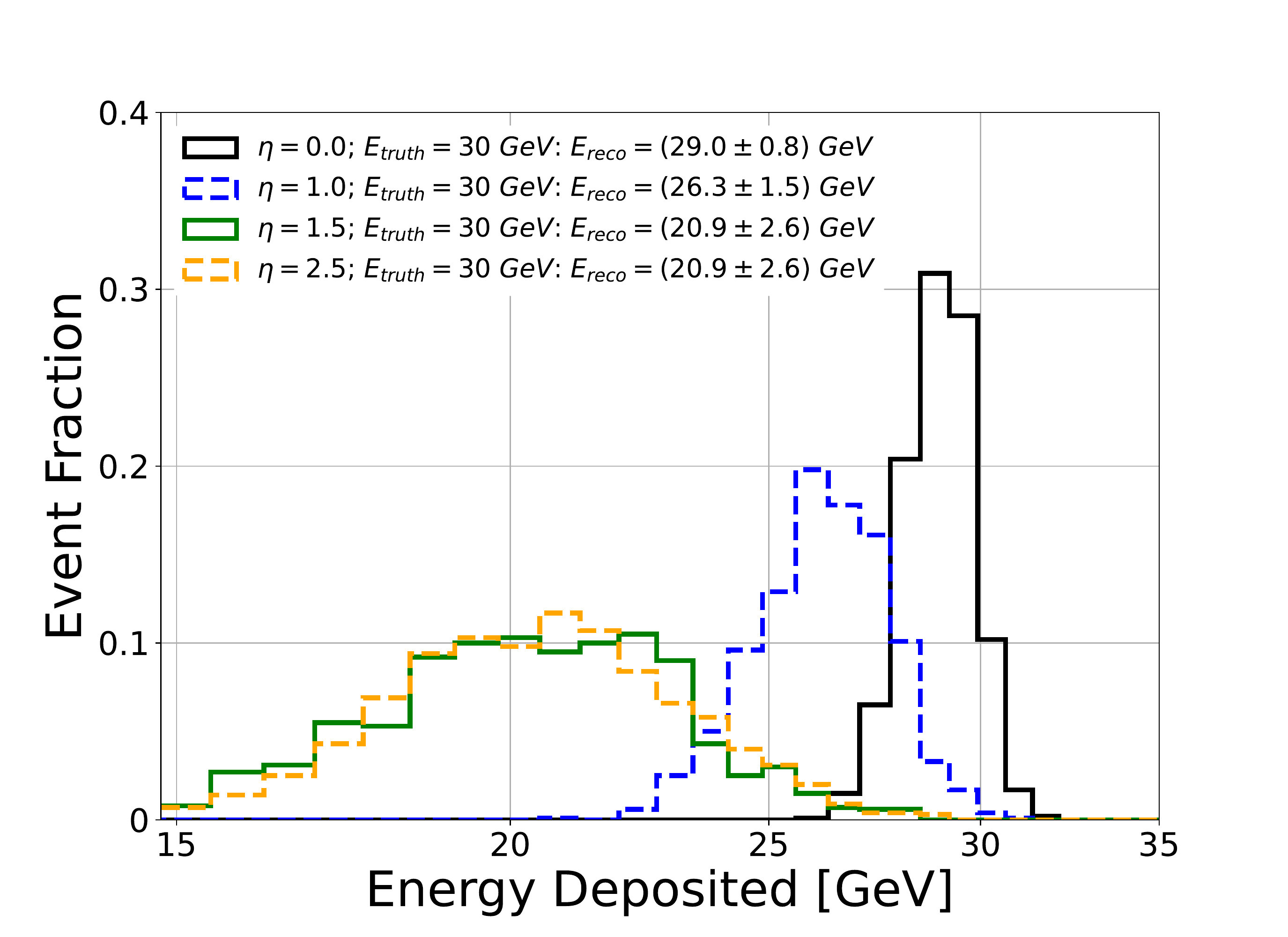}
    \caption{Reconstructed energies of electrons with an energy of 30\,GeV for different directions $\eta$. As the initial electron momentum is directed closer to the beamline, the difference between the reconstructed energy and the intial particle energy increases due to the iron traversed by the electron upstream the calorimeter (dead material).}
    \label{fig:eResponseEtaScan}
\end{figure}

Figure~\ref{fig:eResponseEtaScan} shows the reconstructed energies
of a single electron shot at different initial $\eta$. 
The average reconstructed energy is always lower than the initial particle energies,
with the difference growing with the particle $\eta$. 
This is due to the energy depositions in the iron contained in the ITS upstream the calorimeter, 
in accordance with the material map presented in Fig.~\ref{fig:Rad_Int_length}.

\begin{figure}[!ht]
    \centering
    \includegraphics[width=0.46\textwidth]{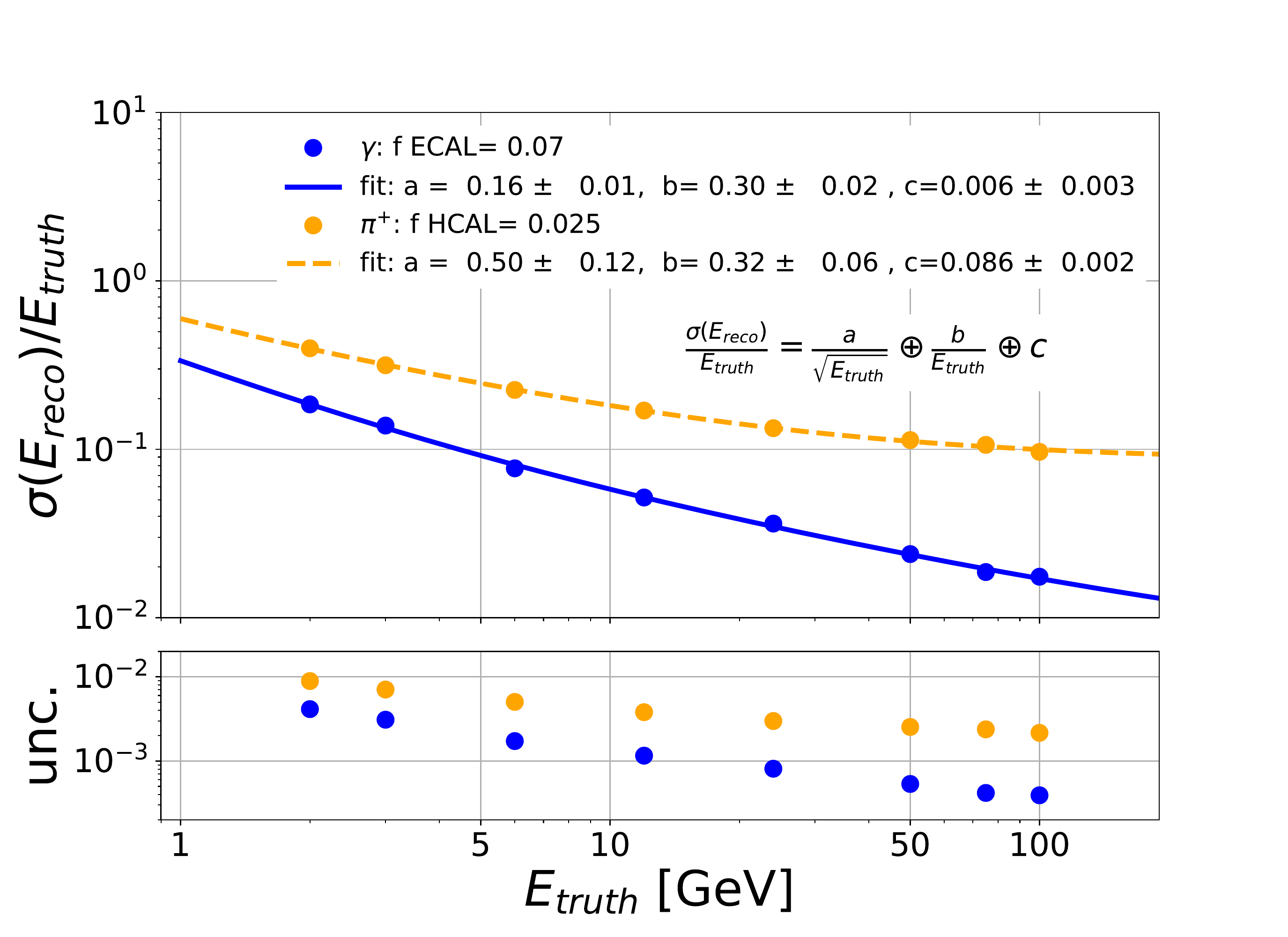}
    \caption{The relative energy resolution $\sigma(E_\mathrm{reco})/E_\mathrm{truth}$ is plotted as a function of $E_\mathrm{truth}$ for eight different truth energy values and fitted with the relative-resolution function, for photon and pion, respectively. The average sampling fraction $f$ for ECAL and HCAL are shown in the legend.}
    \label{fig:eResFits}
\end{figure}

Figure~\ref{fig:eResFits} quantifies the energy resolution as a function of particle energy,
comparing electrons with charged pions.
For each particle type, the relative energy resolution depending
on the particle energy is fitted using least-squares to the following common form of the resolution function:
\begin{equation}
\label{eqn:resol}
\frac{\sigma(E_\mathrm{reco})}{E_\mathrm{truth}} = \frac{a}{\sqrt{E_\mathrm{truth}}} \oplus  \frac{b}{E_\mathrm{truth}} \oplus c
\end{equation}
where the best-fit parameters are provided within the figure. The larger fitted coefficient of the sampling term for hadronic shower compared to electromagnetic is related to the larger value of sampling fraction $f$ configured for the ECAL and HCAL separately (0.07 and 0.025, respectively).
The values of the parameters, appearing in Equation \ref{eqn:resol}, are individually evaluated for photon as $a = 0.16 \pm 0.01$, $b=0.30 \pm 0.02$ and $c = 0.006 \pm 0.003$. The same numbers for charged pions are found to be $a=0.50 \pm 0.12$, $b=0.32\pm0.06$ and $c=0.086\pm0.002$. The noise term is compatible with the input noise values, the sampling term is as expected from the sampling emulation.

\begin{figure}[!h]
    \centering
    \includegraphics[width=0.47\textwidth]{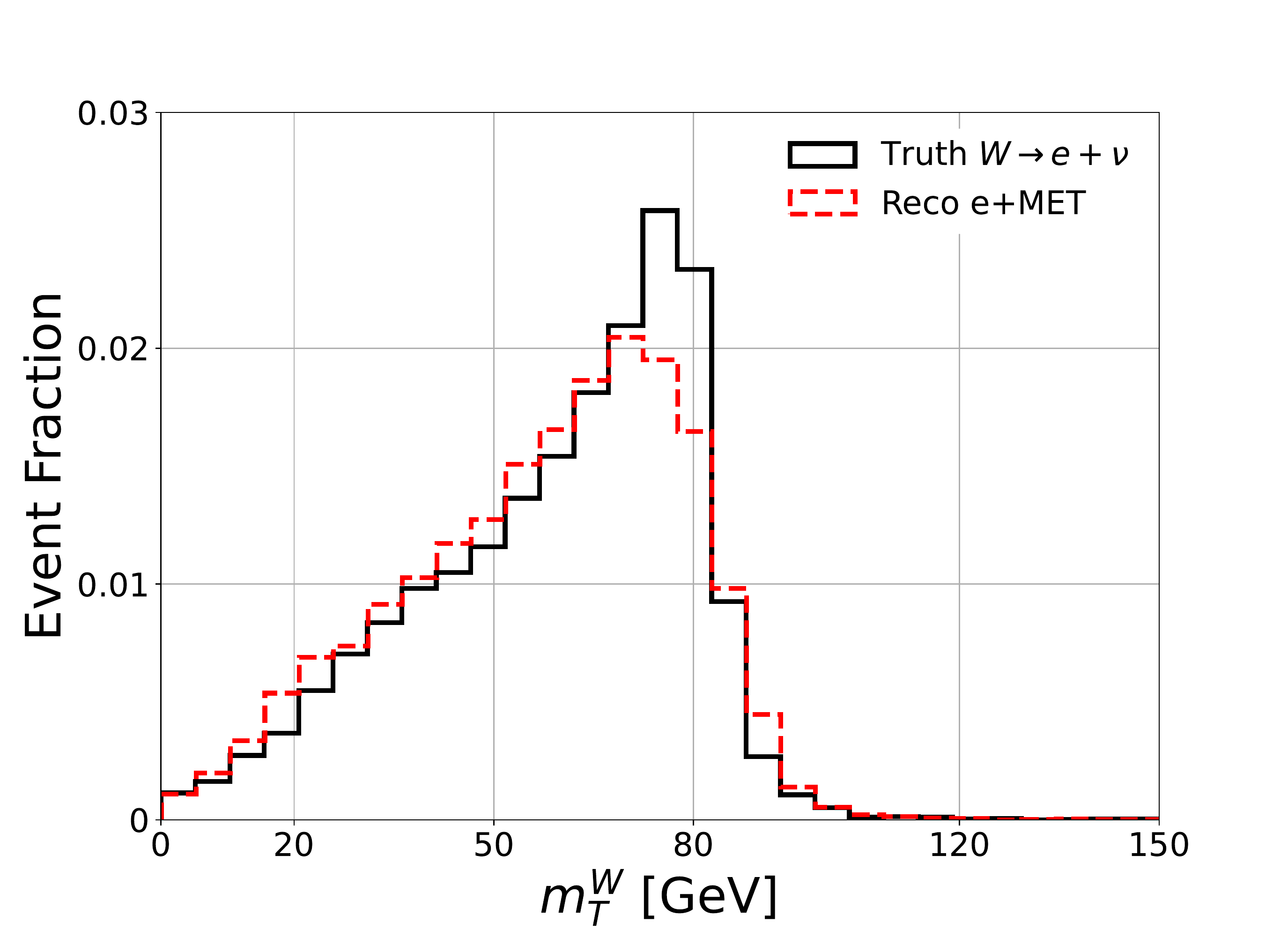}
    \caption{The transverse W mass $m_T^W$ distribution is plotted for leptonically decaying W events. The black curve shows the truth distribution whereas the red curve is obtained from the vector sum of reconstructed lepton momentum and the MET in the event. The peak location of the two distributions are well aligned, demonstrating that the event-level reconstructed MET is trustworthy within the \scd{} framework.}
    \label{fig:mTW}
\end{figure}

The performance of the simulated detector has been so far probed using single particles. To illustrate the detector performance in a more realistic event environment
$pp \rightarrow W \rightarrow e + \nu$ were simulated. The electron is reconstructed using the superclustering algorithm described in Sec. \ref{sec:dataProcessing}, and its energy is calibrated in order to compensate for the the loss due to scattering in the ITS and iron layers upstream the ECAL. The missing transverse momentum (MET) is calculated from the rescaled clusters, as the opposite the vector sum over visible transverse momenta in the whole event. Finally the transverse $W$ mass, $m_T^W$, is computed from the reconstructed $W$ four-momentum and compared with the corresponding truth level distribution in Fig. \ref{fig:mTW}.


\section{Event display}\label{sec:eventDisplay}
\begin{figure*}[!ht]
    \centering
    \subcaptionbox{\label{fig:eventDisplay_ttbar}}[\textwidth]{\hspace{-9mm}\includegraphics[width=0.75\textwidth]{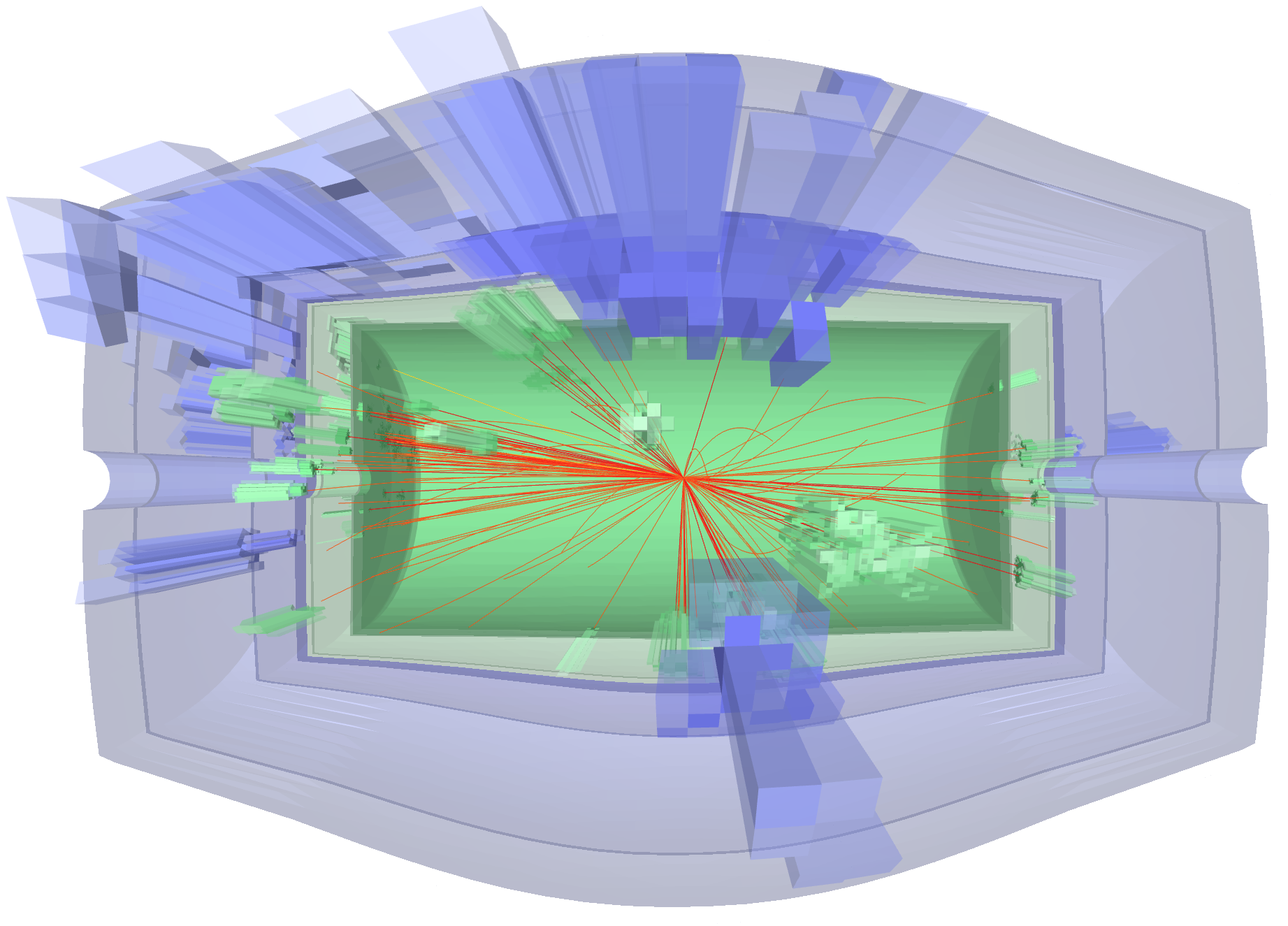}}
    \subcaptionbox{\label{fig:eventDisplay_W}}[\textwidth]{
    \vspace{15mm}\hspace{-40mm} 
    \includegraphics[width=0.75\textwidth]{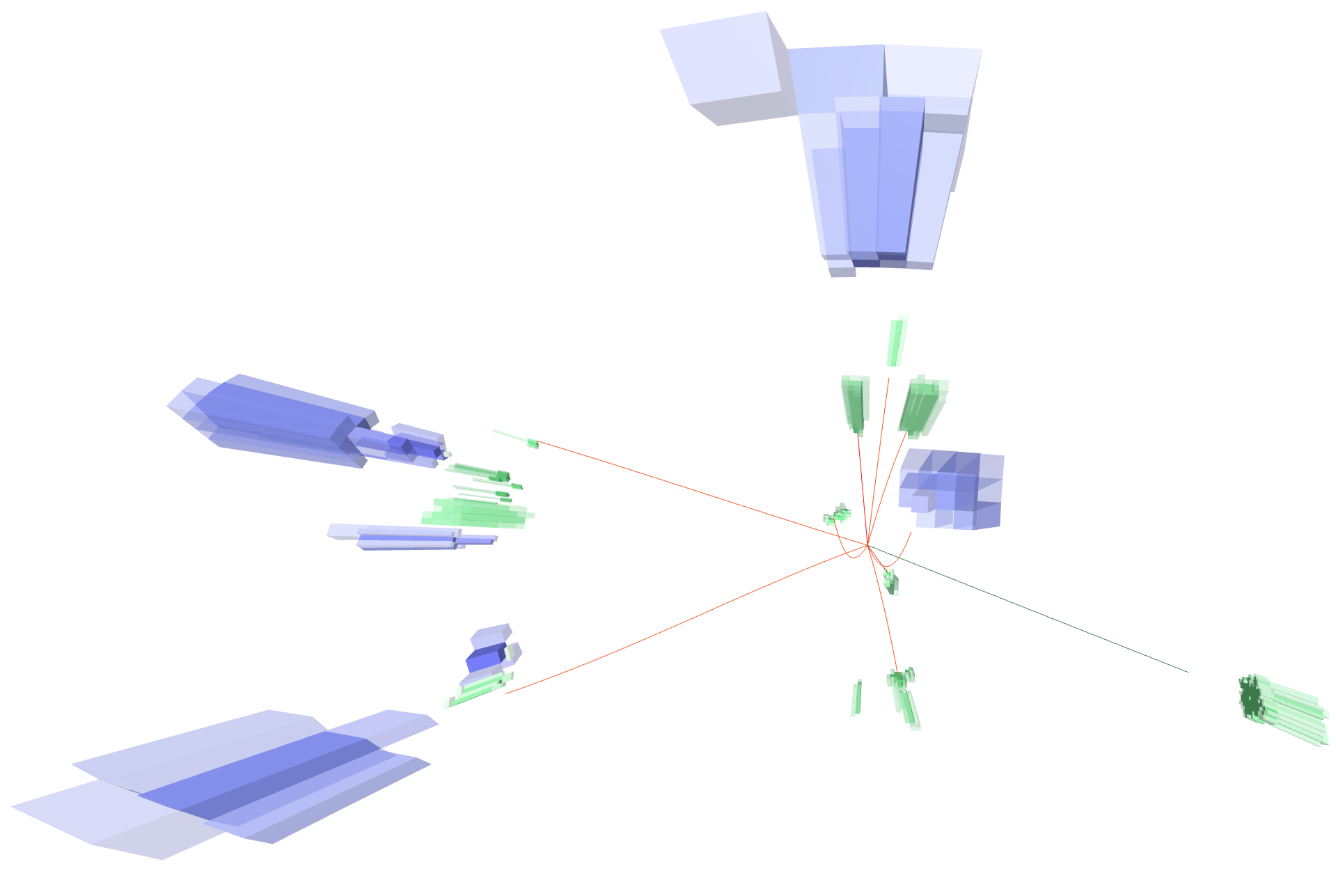}}
    \vspace{10mm}
    \caption{Phoenix event displays configured using the \scd{} detector geometry, showing the charged particle tracks and calorimeter hits generated by (a) $pp\rightarrow t\overline{t}$ and (b) $pp\rightarrow W \rightarrow e\nu$ events simulated with \textsc{Pythia8}. In (a), a cutaway of the \scd{} calorimeter volumes is shown along with the clustered cells, while in (b) only the cells are shown. The electron from the $W$ decay in (b) is indicated by a green line. Both displays are shown in perspective view, such that nearer objects appear larger. Different shades of green and blue represent the different layers of ECAL and HCAL, respectively, while cell opacity is determined by cell signal-to-noise ratio.}
    \label{fig:eventDisplay}
\end{figure*}

Visualization of detector geometry and examples of hits for individual events is important for communicating results, and interpreting downstream tasks such as reconstruction and event selection. The default geometry of the \scd{} detector was ported into the open-source framework Phoenix, chosen for its versatility and user support. An example event display is shown in Fig.~\ref{fig:eventDisplay}.

\section{Conclusion}\label{sec:conclusion}
The  
growing interest in
ML approaches to low-level analysis tasks such as event or jet reconstruction in a realistic detector 
underscores
the importance of leveraging the rich feature space of calorimeter showers 
for improving these tasks. 
Providing an open, configurable, and realistic calorimeter simulation,
\scd{} will facilitate the development of such algorithms and ultimately expand 
the physics reach of current and next-generation collider experiments.
The thorough treatment of particle interactions in \textsc{Geant4} 
and the full-coverage, highly-granular design of \scd{} calorimeter system
enable an accurate representation of the complex data environment 
present in the ATLAS and CMS experiments at the LHC. To quantify this resemblance,
an investigation of the single-particle response characteristics, in terms of
topological clustering performance and energy resolution for electromagnetic and hadronic showers, has been carried out.
Finally, additional aides including data post-processing, 
event visualization, and documentation for \scd{} has been provided to further encourage use.

\section{Acknowledgments}
E.G. would like to thank the United States-Israel BSF , grant NSF-BSF 2020780, for its support. E.D. is supported by the Zuckerman STEM Leadership Program. SG is partially supported by Institute of AI and Beyond for the University of Tokyo.
K.C. and P.R. are supported by NSF award PHY-2111244, and K.C. is also supported by NSF award OAC-1836650.
LH is supported by the Excellence Cluster ORIGINS, which is funded by the Deutsche Forschungsgemeinschaft (DFG, German Research Foundation) under Germany’s Excellence Strategy - EXC-2094-390783311.
\bibliography{Detector}
\bibliographystyle{unsrt}

\end{document}